

\input epsf
%
%
\catcode`\@=11 
\newcount\yearltd\yearltd=\year\advance\yearltd by -1900
%

\def\draftmode{\message{ DRAFTMODE }\def\draftdate{{\rm preliminary draft:
\number\month/\number\day/\number\yearltd\ \ \hourmin}}%
\headline={\hfil\draftdate}\writelabels\baselineskip=20pt plus 2pt minus 2pt
 {\count255=\time\divide\count255 by 60 \xdef\hourmin{\number\count255}
  \multiply\count255 by-60\advance\count255 by\time
  \xdef\hourmin{\hourmin:\ifnum\count255<10 0\fi\the\count255}}}
\def\nolabels{\def\wrlabeL##1{}\def\eqlabeL##1{}\def\reflabeL##1{}}
\def\writelabels{\def\wrlabeL##1{\leavevmode\vadjust{\rlap{\smash%
{\line{{\escapechar=` \hfill\rlap{\sevenrm\hskip.03in\string##1}}}}}}}%
\def\eqlabeL##1{{\escapechar-1\rlap{\sevenrm\hskip.05in\string##1}}}%
\def\reflabeL##1{\noexpand\llap{\noexpand\sevenrm\string\string\string##1}}}
\nolabels
%
\global\newcount\secno \global\secno=0
\global\newcount\meqno \global\meqno=1
\def\newsec#1{\global\advance\secno by1\message{(\the\secno. #1)}
\global\subsecno=0\eqnres@t\noindent{\bf\the\secno ~#1}
\writetoca{{\secsym} {#1}}\par\nobreak\medskip\nobreak}
\def\eqnres@t{\xdef\secsym{\the\secno.}\global\meqno=1\bigbreak\bigskip}
\def\sequentialequations{\def\eqnres@t{\bigbreak}}\xdef\secsym{}
\global\newcount\subsecno \global\subsecno=0
\def\subsec#1{\global\advance\subsecno by1\message{(\secsym\the\subsecno. #1)}
\ifnum\lastpenalty>9000\else\bigbreak\fi
\noindent{\it\secsym\the\subsecno ~#1}\writetoca{\string\quad
{\secsym\the\subsecno.} {#1}}\par\nobreak\medskip\nobreak}
\def\appendix#1{\global\meqno=1\global\subsecno=0\xdef\secsym{\hbox{#1}}
\bigbreak\bigskip\noindent{\bf #1}
\writetoca{{#1}}\par\nobreak\smallskip\nobreak}
%
%
\def\eqnn#1{\xdef #1{(\secsym\the\meqno)}\writedef{#1\leftbracket#1}%
\global\advance\meqno by1\wrlabeL#1}
\def\eqna#1{\xdef #1##1{\hbox{$(\secsym\the\meqno##1)$}}
\writedef{#1\numbersign1\leftbracket#1{\numbersign1}}%
\global\advance\meqno by1\wrlabeL{#1$\{\}$}}
\def\eqn#1#2{\xdef #1{(\secsym\the\meqno)}\writedef{#1\leftbracket#1}%
\global\advance\meqno by1$$#2\eqno#1\eqlabeL#1$$}
%
%
\global\newcount\refno \global\refno=1
\newwrite\rfile
\def\ref{$^{\the\refno}$\nref}
\def\nref#1{\xdef#1{\the\refno.}\writedef{#1\leftbracket#1}%
\ifnum\refno=1\immediate\openout\rfile=refs.tmp\fi
\global\advance\refno by1\chardef\wfile=\rfile\immediate
\write\rfile{\noexpand\item{#1\ }\reflabeL{#1\hskip.31in}\pctsign}\findarg}
\def\findarg#1#{\begingroup\obeylines\newlinechar=`\^^M\pass@rg}
{\obeylines\gdef\pass@rg#1{\writ@line\relax #1^^M\hbox{}^^M}%
\gdef\writ@line#1^^M{\expandafter\toks0\expandafter{\striprel@x #1}%
\edef\next{\the\toks0}\ifx\next\em@rk\let\next=\endgroup\else\ifx\next\empty%
\else\immediate\write\wfile{\the\toks0}\fi\let\next=\writ@line\fi\next\relax}}
\def\striprel@x#1{} \def\em@rk{\hbox{}}
\def\lref{\begingroup\obeylines\lr@f}
\def\lr@f#1#2{\gdef#1{\ref#1{#2}}\endgroup\unskip}
\def\semi{;\hfil\break}
\def\addref#1{\immediate\write\rfile{\noexpand\item{}#1}} 
\def\startrefs#1{\immediate\openout\rfile=refs.tmp\refno=#1}
\def\xref{\expandafter\xr@f}\def\xr@f#1.{#1}
\def\cite{\expandafter\cxr@f}\def\cxr@f#1.{$^{#1}$}
\def\xcite{\expandafter\xcxr@f}\def\xcxr@f#1.{{#1}}
\def\cites#1{\count255=1$^{\r@fs #1{\hbox{}}}$}
\def\r@fs#1{\ifx\und@fined#1\message{reflabel \string#1 is undefined.}%
\nref#1{need to supply reference \string#1.}\fi%
\vphantom{\hphantom{#1}}\edef\next{#1}\ifx\next\em@rk\def\next{}%
\else\ifx\next#1\ifodd\count255\relax\xref#1\count255=0\fi%
\else#1\count255=1\fi\let\next=\r@fs\fi\next}
\newwrite\lfile
{\escapechar-1\xdef\pctsign{\string\%}\xdef\leftbracket{\string\{}
\xdef\rightbracket{\string\}}\xdef\numbersign{\string\#}}

\def\writestop{\def\writestoppt{\immediate\write\lfile{\string\pageno%
\the\pageno\string\startrefs\leftbracket\the\refno\rightbracket%
\string\def\string\secsym\leftbracket\secsym\rightbracket%
\string\secno\the\secno\string\meqno\the\meqno}\immediate\closeout\lfile}}
\def\writestoppt{}\def\writedef#1{}
\def\seclab#1{\xdef #1{\the\secno}\writedef{#1\leftbracket#1}\wrlabeL{#1=#1}}
\def\subseclab#1{\xdef #1{\secsym\the\subsecno}%
\writedef{#1\leftbracket#1}\wrlabeL{#1=#1}}
\newwrite\tfile \def\writetoca#1{}
\def\leaderfill{\leaders\hbox to 1em{\hss.\hss}\hfill}
\def\writetoc{\immediate\openout\tfile=toc.tmp
   \def\writetoca##1{{\edef\next{\write\tfile{\noindent ##1
   \string\leaderfill {\noexpand\number\pageno} \par}}\next}}}

%
\catcode`\@=12 
%
%
\font\abssl=cmsl10 scaled 833
\font\absrm=cmr10 scaled 833 \font\absrms=cmr7 scaled  833
\font\absrmss=cmr5 scaled  833 \font\absi=cmmi10 scaled  833
\font\absis=cmmi7 scaled  833 \font\absiss=cmmi5 scaled  833
\font\abssy=cmsy10 scaled  833 \font\abssys=cmsy7 scaled  833
\font\abssyss=cmsy5 scaled  833 \font\absbf=cmbx10 scaled 833
\skewchar\absi='177 \skewchar\absis='177 \skewchar\absiss='177
\skewchar\abssy='60 \skewchar\abssys='60 \skewchar\abssyss='60
\def\abstractfont{\def\rm{\fam0\absrm}
\textfont0=\absrm \scriptfont0=\absrms \scriptscriptfont0=\absrmss
\textfont1=\absi \scriptfont1=\absis \scriptscriptfont1=\absiss
\textfont2=\abssy \scriptfont2=\abssys \scriptscriptfont2=\abssyss
\textfont\itfam=\absi \def\it{\fam\itfam\absi}
\textfont\slfam=\abssl \def\sl{\fam\slfam\abssl}
\textfont\bffam=\absbf \def\bf{\fam\bffam\absbf}\rm}
\font\ftsl=cmsl10 scaled 833
\font\ftrm=cmr10 scaled 833 \font\ftrms=cmr7 scaled  833
\font\ftrmss=cmr5 scaled  833 \font\fti=cmmi10 scaled  833
\font\ftis=cmmi7 scaled  833 \font\ftiss=cmmi5 scaled  833
\font\ftsy=cmsy10 scaled  833 \font\ftsys=cmsy7 scaled  833
\font\ftsyss=cmsy5 scaled  833 \font\ftbf=cmbx10 scaled 833
\skewchar\fti='177 \skewchar\ftis='177 \skewchar\ftiss='177
\skewchar\ftsy='60 \skewchar\ftsys='60 \skewchar\ftsyss='60
\def\footnotefont{\def\rm{\fam0\ftrm}
\textfont0=\ftrm \scriptfont0=\ftrms \scriptscriptfont0=\ftrmss
\textfont1=\fti \scriptfont1=\ftis \scriptscriptfont1=\ftiss
\textfont2=\ftsy \scriptfont2=\ftsys \scriptscriptfont2=\ftsyss
\textfont\itfam=\fti \def\it{\fam\itfam\fti}%
\textfont\slfam=\ftsl \def\sl{\fam\slfam\ftsl}%
\textfont\bffam=\ftbf \def\bf{\fam\bffam\ftbf}\rm}
\font\ninerm=cmr9 \font\sixrm=cmr6 \font\ninei=cmmi9 \font\sixi=cmmi6
\font\ninesy=cmsy9 \font\sixsy=cmsy6 \font\ninebf=cmbx9
\font\nineit=cmti9 \font\ninesl=cmsl9 \skewchar\ninei='177
\skewchar\sixi='177 \skewchar\ninesy='60 \skewchar\sixsy='60
\def\ninepoint{\def\rm{\fam0\ninerm}
\textfont0=\ninerm \scriptfont0=\sixrm \scriptscriptfont0=\fiverm
\textfont1=\ninei \scriptfont1=\sixi \scriptscriptfont1=\fivei
\textfont2=\ninesy \scriptfont2=\sixsy \scriptscriptfont2=\fivesy
\textfont\itfam=\ninei \def\it{\fam\itfam\nineit}\def\sl{\fam\slfam\ninesl}%
\textfont\bffam=\ninebf \def\bf{\fam\bffam\ninebf}\rm}
%
%

\vsize=7.0truein
\hsize=4.7truein
\baselineskip 12truept plus 0.5truept minus 0.5truept
\hoffset=0.5truein
\voffset=0.5truein

\def\tr{\,{\hbox{tr}}\,}

\def\epm#1#2{\hbox{${+#1}\atop {-#2}$}}

\def\gsim{\mathrel{\rlap{\lower4pt\hbox{\hskip1pt$\sim$}}
    \raise1pt\hbox{$>$}}}         

\def\etal{{\it et al.}}
\def\rhs{right hand side}

\def\frac#1#2{{{#1}\over {#2}}}
\def\half{\hbox{${1\over 2}$}}

\def\smallfrac#1#2{\hbox{${{#1}\over {#2}}$}}

\def\as{\alpha_s}
\def\tr{{\rm tr}}

\def\GeV{{\rm GeV}}
\def\MS{\hbox{$\overline{\rm MS}$}}

\catcode`@=11 
\def\slash#1{\mathord{\mathpalette\c@ncel#1}}
 \def\c@ncel#1#2{\ooalign{$\hfil#1\mkern1mu/\hfil$\crcr$#1#2$}}
\def\lsim{\mathrel{\mathpalette\@versim<}}
\def\gsim{\mathrel{\mathpalette\@versim>}}
 \def\@versim#1#2{\lower0.2ex\vbox{\baselineskip\z@skip\lineskip\z@skip
       \lineskiplimit\z@\ialign{$\m@th#1\hfil##$\crcr#2\crcr\sim\crcr}}}
\catcode`@=12 

\def\PR{{\it Phys.~Rev.~}}
\def\PRL{{\it Phys.~Rev.~Lett.~}}
\def\NP{{\it Nucl.~Phys.~}}
\def\NPBPS{{\it Nucl.~Phys.~B (Proc.~Suppl.)~}}
\def\PL{{\it Phys.~Lett.~}}
\def\PRep{{\it Phys.~Rep.~}}

\def\SJNP{{\it Sov.~Jour.~Nucl.~Phys.~}}
\def\ZP{{\it Zeit.~Phys.~}}
\def\JP{{\it Jour.~Phys.~}}
\def\vol#1{{\bf #1}}\def\vyp#1#2#3{\vol{#1}, #3 (#2)}


\tolerance=10000
\hfuzz=5pt
\pageno=0\nopagenumbers\tolerance=10000\hfuzz=5pt
\line{\hfill {\tt hep-ph/9511330}}
\line{\hfill Edinburgh 95/558}
\vskip 24pt
\centerline{\bf PERTURBATIVE EVOLUTION OF}
\centerline{\bf POLARIZED STRUCTURE FUNCTIONS}
\vskip 36pt\centerline{Richard D. Ball\footnote*{\footnotefont
CERN Fellow and Royal Society University Research Fellow}}
\vskip 12pt
\centerline{\it Theory Division, CERN}
\centerline{\it CH-1211 Gen\`eve 23, Switzerland}
\centerline{and}
\centerline{\it Department of Physics and Astronomy}
\centerline{\it University of Edinburgh, EH9 3JZ, Scotland}
\vskip 48pt
{\narrower\baselineskip 10pt
\centerline{\bf Abstract}
\medskip\noindent
We review the perturbative evolution of the polarized structure functions
$g_1$ and their associated parton distribution functions, with particular
emphasis on the anomalous coupling of the first moment of the polarized
gluon distribution.  We also describe the small $x$ behaviour of polarized
parton distributions, contrasting it with that of the unpolarized
distributions. We then explain how this theoretical analysis affects the
extraction of the singlet axial charge from experimental data on $g_1$,
and show that it may be possible to use such data to infer the
existence of polarized gluons in the nucleon.
\smallskip}
\bigskip
\centerline{Invited talk at the}
\centerline{\it International School of Nucleon Spin Structure}
\centerline{Erice, August 1995}
\medskip
\centerline{\it to be published in the proceedings}
\vskip 55pt
\line{November 1995\hfill}
\eject
\footline={\hss\tenrm\folio\hss}
\centerline{\bf PERTURBATIVE EVOLUTION OF}
\centerline{\bf POLARIZED STRUCTURE FUNCTIONS}
\bigskip\bigskip
{\ninepoint
\centerline{R.D.~BALL}
\smallskip
\centerline{\it Theory Division, CERN, CH-1211 Gen\`eve 23, Switzerland}
\centerline{and}
\centerline{\it Department of Physics and Astronomy,}
\centerline{\it University of Edinburgh, EH9 3JZ, Scotland}
}
\bigskip
{\abstractfont\baselineskip 9 pt
\advance\leftskip by 36truept\advance\rightskip by 36truept\noindent
We review the perturbative evolution of the polarized structure functions
$g_1$ and their associated parton distribution functions, with particular
emphasis on the anomalous coupling of the first moment of the polarized
gluon distribution.  We also describe the small $x$ behaviour of polarized
parton distributions, contrasting it with that of the unpolarized
distributions. We then explain how this theoretical analysis affects the
extraction of the singlet axial charge from experimental data on $g_1$,
and show that it may be possible to use such data to infer the
existence of polarized gluons in the nucleon.
\smallskip}

\baselineskip 12pt plus 0.5pt minus 0.5pt
\bigskip\bigskip
\goodbreak


\nref\EMC{EMC Collaboration, J.~Ashman \etal, \NP\vyp{B328}{1989}{1}.}
\nref\AlRo{G.~Altarelli and G.~G.~Ross, \PL\vyp{B212}{1988}{391}.}
\nref\CCM{R.D.~Carlitz, J.C.~Collins and A.H.~Mueller,
                                       \PL\vyp{B214}{1988}{229}.}
\nref\AlLa{G.~Altarelli and B.~Lampe, \ZP\vyp{C47}{1990}{315}.}
\nref\guidoerice{G.~Altarelli, in ``The Challenging Questions'',
 Proc. of the 1989 Erice School, A.~Zichichi, ed. (Plenum, New York, 1990).}
\nref\BoQiu{G.T.~Bodwin and J.~Qiu, \PR\vyp{D41}{1990}{2755}.}
\nref\Vogel{W.~Vogelsang, \ZP\vyp{C50}{1991}{275}.}
\nref\SMCp{SMC Collaboration, D.~Adams \etal, \PL\vyp{B329}{1994}{399}.}
\nref\SMCd{SMC Collaboration, D.~Adams \etal, \PL\vyp{B357}{1995}{248}.}
\nref\SLACp{E143 Collaboration, K.~Abe \etal, \PRL\vyp{74}{1995}{346}.}
\nref\SLACd{E143 Collaboration, K.~Abe \etal, \PRL\vyp{75}{1995}{25}.}
\nref\AlRi{G.~Altarelli and G.~Ridolfi, \NPBPS\vyp{39B}{1995}{106}.}
\nref\stefrev{S.~Forte, {\tt hep-ph/9409416}, in ``Radiative
   Corrections: Status and Outlook'', B.~F.~L.~Ward, ed. (World Scientific,
   Singapore, 1995).}
\nref\BFR{R.D.~Ball, S.~Forte and G.~Ridolfi, \NP\vyp{B444}{1995}{287}.}

\noindent
The publication of the EMC results\cite\EMC\ for the polarized proton
structure function $g_1^p$ has been directly responsible for a renewed
interest in polarized deep inelastic scattering among the theoretical
community. In particular the implication that in the naive parton
model the total helicity carried by quarks and antiquarks in the
proton was consistent with zero led to a rexamination of the role
played by the axial anomaly and polarized gluons in the perturbative
evolution of the first moment of polarized structure
functions.\cites{\AlRo-\Vogel} Since then, more precise
measurements\cites{\SMCp-\SLACd} of both $g_1^p$ and $g_1^d$ over a
wider range of both $x$ and $Q^2$ have been made, and from them
seemingly very precise values of the first moments
deduced.\cites{\AlRi,\stefrev} These first moments are generally
obtained by extrapolating the experimental data to a common scale
(which is done in practice by assuming that the asymmetries
measured in the experiments are $Q^2$ independent), and further by
extrapolating from the measured region to small $x$ (using Regge
behaviour, and assuming that the small $x$ contribution is then $Q^2$
independent). However if the polarized gluon distribution were large,
both of these approximations could turn out to be very poor,\cite\BFR\
because of the anomalously large coupling of polarized gluons to the
first moment of $g_1$. It thus
becomes necessary to examine in detail the theoretical errors implicit
in our present ignorance of the size of the polarized gluon distribution, and
conversely whether the $x$ and $Q^2$ dependence of existing or future
structure function data may be used to infer the existence of a large
gluonic contribution to the nucleon spin.

\newsec{Polarized Partons}
\nref\guidorev{G.~Altarelli, \PRep\vyp{81}{1982}{1}.}
\nref\AlPa{G.~Altarelli and G.~Parisi, \NP\vyp{B126}{1977}{298}.}
\nref\gott{R.D.~Ball and S.~Forte, \NP\vyp{B425}{1994}{516}.}
\nref\Kod{J.~Kodaira \etal, \PR\vyp{D20}{1979}{627};
                                \NP\vyp{B159}{1979}{99}\semi
               J.~Kodaira, \NP\vyp{B165}{1979}{129}.}

We begin by reviewing the relation between polarized structure
functions and parton densities in the parton model, and its relation to
the operator product expansion and renormalization group
approach. In the parton model\cite\guidorev\ the polarized structure
function $g_1$ is decomposed in terms of polarized quark and gluon
distributions $\Delta q_i$ and $\Delta g$ according to
\eqn\gone
{\eqalign{g_1(x,Q^2)=
&\frac{\langle e^2\rangle}{2}\int_x^1\! \frac{dy}{y}\,\Big\{
      C_{\rm NS}(\smallfrac{x}{y},\as(t))\Delta q_{\rm NS}(y,t)\cr
      +&C_{\rm S}(\smallfrac{x}{y},\as(t))\Delta q_{\rm S}(y,t)
      +C_g(\smallfrac{x}{y},\as(t))\Delta g(y,t)\Big\}+O\big(1/Q^2\big),}}
where $t\equiv\ln\frac{Q^2}{\Lambda^2}$, the various coefficient functions
$C(x,\as)$ are directly related to hard
cross-sections calculable in perturbative QCD, and $\Delta q_{\rm NS}$ and
$\Delta q_{\rm S}$ are respectively the nonsinglet and singlet quark
distributions:
\eqn\qsing{\eqalign
{\Delta q_{\rm NS}(x,t)&\equiv\sum_{i=1}^{n_f}
\smallfrac{e^2_i-\langle e^2\rangle}{\langle e^2\rangle}
(\Delta q_i(x,t)+\Delta\bar q_i(x,t)),\cr
\Delta q_{\rm S}(x,t)&\equiv \sum_{i=1}^{n_f}
(\Delta q_i(x,t)+\Delta\bar q_i(x,t)),}
}
where $n_f$ is the number of active flavours, each with electric charge
$e_i$, and $\langle e^2\rangle\equiv\sum e_i^2/n_f$. Although they are
themselves intrinsically nonperturbative, the perturbative part of
the $x$ and $t$ dependence of the polarized quark and
gluon distributions is given by Altarelli-Parisi equations:\cite\AlPa\
the nonsinglet quark evolves independently as
\eqn\apqns
{\frac{d}{dt}\Delta q_{\rm NS}(x,t)
=\frac{\as(t)}{2\pi}\int_x^1\! \frac{dy}{y}\,
P_{qq}^{\rm NS}(\smallfrac{x}{y},\as(t))\Delta q_{\rm NS}(y,t),}
while the singlet quark and the gluon mix according to
\eqn\aps{\eqalign{
\frac{d}{dt}\Delta q_{\rm S}(x,t)
&=\frac{\as(t)}{2\pi}\int_x^1\! \frac{dy}{y}\,\left[
P_{qq}^{\rm S}(\smallfrac{x}{y},\as(t))\Delta q_{\rm S}(y,t)
          + P_{qg}(\smallfrac{x}{y},\as(t))\Delta g(y,t)\right],\cr
\frac{d}{dt}\Delta g(x,t)
&=\frac{\as(t)}{2\pi}\int_x^1\! \frac{dy}{y}\,\left[
P_{gq}(\smallfrac{x}{y},\as(t))\Delta q_{\rm S}(y,t)
                 + P_{gg}(\smallfrac{x}{y},\as(t))\Delta g(y,t)\right].}}
The splitting functions $P(x,\as)$ are again computable perturbatively
in terms of hard cross-sections.  In the naive parton model
$C_{\rm NS}=C_{\rm S}=\delta(1-x)$, $C_g=0$, so the gluons decouple.
In LO perturbation theory the gluons couple through the singlet evolution
equations \aps, while at NLO they also couple directly. Heavy quark
contributions are generated radiatively as thresholds are crossed. All
this works in just the same way as in the unpolarized case.

Taking the Mellin transform of eq.\gone\ gives the leading twist
component of the operator product expansion of the moments of $g_1$:
\eqn\gonemom{\eqalign{
\Gamma_N(Q^2)&\equiv\hbox{$\int_0^1$} \,dx x^{N-1} g_1(x,Q^2)\cr
&=\smallfrac{\langle e^2\rangle}{2}\left[
C^{\rm NS}_N\Delta q^{\rm NS}_N
+ C^{\rm S}_N\Delta q^{\rm S}_N
+ C^g_N\Delta g_N\right] + O\big(1/Q^2\big).\cr}}
Here $C_N(\as)\equiv\int_0^1 \,dx x^{N-1}C(x,\as)$ are the Wilson
coefficients and $\Delta q_N(t)\equiv\int_0^1 \,dx x^{N-1}\Delta
q(x,t)$ may, for some values of $N$, be related to forward matrix elements of
local operators. In the unpolarized case moments of the distributions
$q_i+\bar q_i$ and $g$ correspond to matrix elements of local twist
two operators for
$N=2,4,\ldots$, while for $q_i-\bar q_i$ suitable local operators exist for
$N=1,3,\ldots$: all the other moments are well defined, but can only
be obtained by analytic continuation in $N$ (a good
example\cite\gott\ being the Gottried sum $q_1^{\rm NS}$). Here however
the situation is reversed:\cite\Kod
\eqn\matel{\eqalign{
s_\mu p_{\nu_1}\ldots p_{\nu_{N-1}}\Delta q^{\rm NS}_N(t)&
=\langle p,s\vert{\cal O}^{{\rm NS},N}_{\mu,\nu_1,\ldots,\nu_{N-1}}
\vert p,s\rangle_t,
\qquad N=1,3,5,\ldots\cr
s_\mu p_{\nu_1}\ldots p_{\nu_{N-1}}\Delta q^{\rm S}_N(t)&
=\langle p,s\vert{\cal O}^{{\rm S},N}_{\mu,\nu_1,\ldots,\nu_{N-1}}
\vert p,s\rangle_t,
\qquad N=3,5,7,\ldots\cr
s_\mu p_{\nu_1}\ldots p_{\nu_{N-1}}\Delta g_N(t)&
=\langle p,s\vert{\cal O}^{{\rm g},N}_{\mu,\nu_1,\ldots,\nu_{N-1}}
\vert p,s\rangle_t,
\qquad N=3,5,7,\ldots,\cr}}
where $\vert p,s\rangle$ is some hadronic state carrying momentum $p$,
with polarization vector $s_\mu$,
and the twist two local operators ${\cal O}^{\rm g}_N$ are purely gluonic.
Although there exist local operators for even $N=2,4,\ldots$,
these have opposite charge conjugation and thus correspond to moments
of the valence distributions $q_i-\bar q_i$. Again all other moments
are well defined, but can only be obtained by analytic continuation.
They are necessarily gauge invariant since the matrix elements \matel\
are gauge invariant, but will in general be scheme dependent. The case $N=1$
in the singlet channel is peculiar, in that there is just one local
operator, the axial singlet current: the identification of matrix
elements of this operator with the first moments $\Delta q^{\rm S}_1$
and $\Delta g_1$ is then rather subtle and will be discussed in the
next section.

Taking moments of the Altarelli-Parisi equations \apqns,\aps\ yields
the renormalization group equations for the matrix elements \matel:
\eqn\rengrp{\eqalign
{\frac{d}{dt}\Delta q^{\rm NS}_N(t)
&= \gamma^{qq,{\rm NS}}_{N}\big(\as(t)\big)
\Delta q^{\rm NS}_N(t)\cr
\frac{d}{dt}\pmatrix{\Delta q^{\rm S}_N(t)\cr \Delta g_N(t)\cr}&=
\pmatrix{\gamma^{qq,{\rm S}}_{N}\big(\as(t)\big)&
         \gamma^{qg}_N\big(\as(t)\big)\cr
         \gamma^{gq}_N\big(\as(t)\big)&
         \gamma^{gg}_N\big(\as(t)\big)\cr}
\pmatrix{\Delta q^{\rm S}_N(t)\cr \Delta g_N(t)\cr},\cr}}
where $\gamma_N(\as)\equiv\smallfrac{\as}{2\pi}
\int_0^1 \,dx x^{N-1}P(x,\as)$ are the
anomalous dimensions of the various local operators.

\newsec{First Moments}
\nref\Bj{J.D.~Bjorken, \PR\vyp{148}{1966}{1467}.}
\nref\EJ{J.~Ellis and R.L.~Jaffe, \PR\vyp{D9}{1974}{1444}.}
\nref\myphysrep{For a general review see R.D.~Ball, \PRep\vyp{182}{1989}{1}.}
\nref\Larin{S.A.~Larin, \PL\vyp{B334}{1994}{192}.}
\nref\ShVe{G.M.~Shore \etal,
                           \PL\vyp{B244}{1990}{75}; \NP\vyp{B381}{1992}{23}.}
\nref\Forte{S.~Forte, \NP\vyp{B331}{1990}{1}.}
\nref\AdBa{S.~Adler and W.~Bardeen, \PR\vyp{182}{1969}{1517}.}
\nref\GoLa{S.G.~Gorishny and S.A.~Larin, \PL\vyp{B172}{1986}{109}.}
\nref\LaVe{S.A.~Larin and J.A.M.~Vermaseren, \PL\vyp{B259}{1991}{345}.}

Due to interest in the Bjorken\cite\Bj\ and Ellis-Jaffe\cite\EJ\
sum rules, much of the literature on $g_1(x,Q^2)$ has
focussed on its first moment $\Gamma_1(Q^2)$. The proper
interpretation\cites{\AlRo-\guidoerice} of the singlet first
moments is complicated by the presence in this channel of the axial
anomaly.\cite\myphysrep\

Consider firstly the renormalization group equations \rengrp\ with
$N=1$. While some elements of the matrix of anomalous dimensions are
nonvanishing at LO (one loop), this dependence turns out to be
trivial, since the eigenvectors of the evolution
\eqn\evecs{\Delta q^{\rm NS}_1,\qquad\Delta q^{\rm S}_1\qquad{\rm and}
\qquad a_0\equiv\Delta q^{\rm S}_1-n_f\smallfrac{\as}{2\pi}\Delta g_1,}
only evolve at NLO (two loops).\cite\AlRo\ It follows that there must exist
factorization schemes in which both $\Delta q^{\rm NS}_1$ and
$\Delta q^{\rm S}_1$ do not evolve at all, since multiplicative
renormalizations which only begin at NLO can always be removed by a
change of scheme. Thus in such schemes
\eqn\evecevol
{\smallfrac{d}{dt}\Delta q^{\rm NS}_1=0,
\qquad\smallfrac{d}{dt}\Delta q^{\rm S}_1=0\qquad{\rm and}
\qquad \smallfrac{d}{dt}a_0=\gamma_s a_0,}
where $\gamma_s=-n_f(\as^2/2\pi^2)+\cdots$ has been calculated at
two\cite\Kod\ and three\cite\Larin\ loops.

In the nonsinglet sector such schemes are essential in order to make
the usual identification \matel\ of $\Delta q^{\rm NS}_1$ with forward
matrix elements of (partially) conserved nonsinglet axial currents,
i.e. with the nonsinglet axial charges. These may then be determined
(assuming exact flavor symmetry) from weak decays of hyperons: below
the charm threshold
\eqn\nsaxch{\Delta q^{\rm NS}_1=\smallfrac{3}{4}g_A
+ \smallfrac{1}{4}a_8,}
where $s_\mu g_A = \langle p,s\vert (J^5_\mu)_3\vert p,s\rangle$, etc.
Above heavy quark thresholds
nonsinglet charges $a_{15}$, $a_{24}$, etc. must also be added.
The nonsinglet
combinations $g_A=\Delta u-\Delta d$ and $a_8=\Delta u+\Delta d-
2\Delta s$, etc. of the quark contributions to the spin of the hadron
are then both well defined and scale independent (above threshold).
The Bjorken sum rule\cite\Bj\ follows immediately from eqn.\gonemom.

Similarly in the singlet sector the conservation of $\Delta q^{\rm S}_1$
makes it the natural candidate\cites{\AlRo-\guidorev} for the
singlet quark contribution
$\Delta u+\Delta d+\Delta s +\cdots$ to the hadron spin. Individual
quark contributions may then be disentangled, and in particular
the Zweig rule
$0\simeq\Delta s\simeq\Delta c\simeq\cdots$ acquires a scale
independent meaning.\cite\ShVe\ In fact although there is no
local gauge invariant conserved current in the axial singlet channel,
$\Delta q^{\rm S}_1$ may still be formally identified with
singlet quark helicity.\cite\Forte\ The other evolution eigenvector
$a_0$ may then be identified with forward matrix elements
$\langle p,s\vert j^5_\mu\vert p,s\rangle_t = s_\mu a_0(t)$ of
the singlet axial current $j^5_\mu$, whose conservation
is violated at one loop by the (purely gluonic) axial anomaly,
\eqn\anomaly{\partial_\mu j^5_\mu = n_f\smallfrac{\as}{2\pi}
\epsilon_{\mu\nu\rho\sigma}\tr G_{\mu\nu}G_{\rho\sigma},}
which is thus directly responsible for the NLO evolution
\evecevol\ of $a_0$.
Since the anomaly \anomaly\ is unaffected by higher order perturbative
corrections,\cite\AdBa\ it is possible to find schemes in which both the
decomposition \evecs\ of $a_0$ and its evolution \evecevol\ hold to
all orders in perturbation theory:\cite\AlRo\ in such schemes
\eqn\adab{
-n_f\smallfrac{\as}{2\pi}\gamma^{gq}_{1}(\as)=\gamma_s(\as),\qquad
\gamma^{gg}_{1}(\as)=\gamma_s(\as)+\beta(\as),}
where $\beta(\as)\equiv d\ln\as/dt$.

Now consider the Wilson coefficients in the operator product expansion
\gonemom. When these are calculated in a factorization scheme in which
first moments evolve according to \evecevol, their first moments are
\eqn\cofm{C^{\rm NS}_1=1-\smallfrac{\as}{\pi}+\cdots,\qquad
C^{\rm S}_1=1-\smallfrac{\as}{\pi}+\cdots,\qquad
C^{\rm g}_1=-n_f\smallfrac{\as}{2\pi}+\cdots,}
at NLO (in fact $C^{\rm NS}_1$ is known to\cite\GoLa\ $O(\as^3)$,
$C^{\rm S}_1$ to\cite\LaVe\ $O(\as^2)$). When combined with \evecs,
this implies that when $N=1$ there are indeed only two terms on the
\rhs\ of \gonemom:
\eqn\gfmao{\Gamma_1(Q^2)=
\smallfrac{\langle e^2\rangle}{2}
\left[C^{\rm NS}_1(\as)
\Delta q^{\rm NS}_1(t) +C^{\rm S}_1(\as)a_0(t)\right],}
in accordance with the fact that there is only one local gauge
invariant singlet operator with twist two and spin one, the axial
singlet current $j^5_\mu$, which must then be multiplicatively renormalized.
Although \gonemom,\cofm\ and \evecs\ only imply \gfmao\ at NLO,
in schemes in which the Adler-Bardeen theorem is satisfied it must
be true to all orders in perturbation theory: in such schemes
\eqn\sgab{C^g_1(\as)=-n_f\smallfrac{\as}{2\pi}C^{\rm S}_1(\as).}

In practice all of these results may be obtained by regularizing infrared
collinear divergences by putting external particles
off-shell.\cite\Kod\ Alternatively the infrared divergences may
regulated by giving the quarks a mass;\cite\AlRo\ this has the
disadvantage that chiral symmetry is broken, so a finite
renormalization must be performed to ensure that the quark helicities
$\Delta q^{\rm NS}_1$ and $\Delta q^{\rm S}_1$ remain scale
independent. More seriously, if dimensional regularization alone is
used, with massless quarks and all external particles on-shell,
$C^g_1$ vanishes, so $\Delta q^{\rm S}_1$ is identified with the
axial singlet charge and thus can no longer be related to quark
helicity. Such factorization schemes are inappropriate because
soft contributions are included in the coefficient function, rather
than being factorized into the parton densities.\cites{\CCM,\BoQiu,\Vogel}

\nref\FoSh{S.~Forte and E.V.~Shuryak, \NP\vyp{B357}{1991}{153}.}
\nref\ss{R.D.~Ball. \PL\vyp{B266}{1991}{473}.}
\nref\NSV{S.~Narison, G.~Shore and G.~Veneziano, \NP\vyp{B433}{1995}{209}.}

Finally we consider the implications of this analysis. Combining
\evecs\ and \evecevol, we see that $\Delta g_1(t)\sim t$ as
$t\to\infty$, so asymptotically the gluonic contribution to $a_0(t)$
does not decouple, but may be as large as the quark
contribution.\cite\AlRo\
Thus even if $\Delta q^{\rm S}_1\sim a_8$ as suggested by the Zweig
rule, $a_0(t)$ may still be very different from $a_8$, and the
Ellis-Jaffe sum rule\cite\EJ\ may fail.

To actually explain why experimentally\cite\EMC\ $a_0(t)$ is small in
the perturbative region we would nevertheless require some nonperturbative
mechanism. Various attempts at natural expanantions
exist, in which either $\Delta q^{\rm S}_1$ is suppressed by
instantons,\cite\FoSh\ and thus the Zweig rule is strongly violated,
or else the Zweig rule holds but $a_0(t)$ is suppressed, either by
strong evolution at low scales,\cite\ss\ or due to the smallness of
the first derivative of the topological susceptibility.\cite\NSV\ The
latter explanations are both target independent, so might be tested by
measuring $g_1$ for other targets such as the photon. Meanwhile, it
would be useful to have some way of determining $\Delta q^{\rm S}$
or $\Delta g$ independently.

Since the gluonic contribution to the first moment of $g_1$ is
effectively LO, it might be reasonable to expect that if $\Delta g$ were
indeed large the scale dependence of $g_1(x,Q^2)$ might also be anomalously
large in certain regions of $x$.\cite\BFR\ If this were so it might be
possible conversely to determine $\Delta g$ by studying
scaling violations of $g_1$. We will return to this idea in section~4, but
first will discuss another way of arriving at the same conclusion.

\newsec{Small $x$ Evolution}

\nref\DGPTWZ{A.~De~Rujula \etal, \PR\vyp{D10}{1974}{1649}.}
\nref\das{R.D.~Ball and S.~Forte, \PL\vyp{B335}{1994}{77};
                                      \vyp{B336}{1994}{77}.}
\nref\mont{R.D.~Ball and S.~Forte, \NPBPS\vyp{39B,C}{1995}{25}.}
\nref\Gross{D.~Gross in the proceedings of the XVII Intern. Conf. on
                       High Energy Physics, London, 1974
                       (published by SRC, Rutherford Lab.).}
\nref\BFKL{L.N.~Lipatov et al, \SJNP\vyp{23}{1976}{338};
       {\it Sov. Phys. JETP~}\vyp{44}{1976}{443};\vyp{45}{1977}{199};
       \SJNP\vyp{28}{1978}{822}.}
\nref\CCFM{S.~Catani \etal, \PL\vyp{B234}{1990}{339};
                     \NP\vyp{B336}{1990}{18}; \NP\vyp{B361}{1991}{645}.}
\nref\Jaro{T.~Jaroszewicz, \PL\vyp{B116}{1982}{291}.}
\nref\summing{R.D.~Ball and S.~Forte, \PL\vyp{B351}{1995}{313};
                                            \vyp{B358}{1995}{365}.}
\nref\KiLi{R.~Kirschner \etal, \PR\vyp{D26}{1982}{1202};
                                               \NP\vyp{B213}{1983}{122}.}
\nref\EMR{B.I.~Ermolaev S.I.~Manayenkov and M.G.~Ryskin,
                       {\tt hep-ph/9502262}.}
\nref\MevN{R.~Mertig and W.L.~van~Neerven, {\tt hep-ph/9506451}
(revised November 1995).}
\nref\BlVo{J.~Bl\"umlein and A.~Vogt, {\tt hep-ph/9510410}.}
\nref\ZivN{E.B.~Zijlstra \etal,
                   \NP\vyp{B383}{1992}{385}; \vyp{B417}{1994}{61}.}
\nref\Heim{R.L.~Heimann, \NP\vyp{B64}{1973}{429}.}
\nref\Kuti{J.~Kuti, these proceedings.}
\nref\AhRo{M.A.~Ahmed and G.G.~Ross, \PL\vyp{B56}{1975}{385}.}
\nref\EiSo{M.B.~Einhorn and J.~Soffer, \NP\vyp{B74}{1986}{714}.}
\nref\Berera{A.~Berera, \PL\vyp{B293}{1992}{445}.}
\nref\pg{R.D.~Ball, S.~Forte and G.~Ridolfi, {\tt hep-ph/9510449}.}
\nref\BER{J.~Bartels B.I.~Ermolaev and M.G.~Ryskin, {\tt hep-ph/9507271}.}
\nref\FiKi{M.~Fippel and R.~Kirschner, \JP\vyp{G17}{1991}{421}.}

The behaviour of structure functions at small $x$ is governed both by
the qualitative form of the non-perturbative input at some scale
$t_0\equiv \ln Q_0^2/\Lambda^2$, given in principle by Regge
theory, and the structure of the perturbative evolution to
$Q^2>Q_0^2$, which can then often overwhelm it. Indeed the form and
evolution of structure functions at small $x$ depends on the
outcome of a competition between Regge behaviour, LO evolution and
higher order perturbative evolution. We will begin by
describing how this works for unpolarized distributions, and then
consider the polarized distributions.

According to Regge theory the behaviour as $x\to 0$ of the unpolarized singlet
distributions is controlled by the pomeron intercept, so
$xq_{\rm S}(x,t_0)\sim xg(x,t_0)\sim x^{-\lambda}$, with
$\lambda\simeq 0.08$. However this essentially
flat (or `soft') behaviour is substantially modified by perturbative
evolution, determined by the leading singularity which for singlet
anomalous dimensions is
at $N=1$. At one loop, $\gamma^{gg}_N\sim\gamma^{gq}_N\sim \as/(N-1)$
as $N\to 1$, while $\gamma^{qg}_N$ and $\gamma^{qq}_N$ are both
regular. It follows that in the double limit $x\to 0$ and $t\to\infty$
the gluon distribution grows in a precisely determined way, faster than
any power of $\ln 1/x$ but slower than any power of $x$:\cite\DGPTWZ
\eqn\dasg{ xg(x,t)\sim {\cal N}\sigma^{-1/2}
\exp({2\gamma\sigma-\delta\zeta}),}
where if $\xi\equiv\ln\smallfrac{x_0}{x}$,
$\zeta\equiv\ln\smallfrac{\alpha_s(Q_0^2)}{\alpha_s(Q^2)}\sim
\ln\smallfrac{t}{t_0}$, $\sigma\equiv\sqrt{\xi\zeta}$, and $\gamma$
and $\delta$ are (known) numerical constants.  This growth
in $xg$ drives a similar `double scaling' behaviour in $xq_{\rm S}$, which
has recently been confirmed by measurements of $F_2^p$ at small $x$
and large $t$ at HERA.\cite\das\

Two loop corrections to double scaling are small\cite\mont\
essentially  because the
singularity in the two loop singlet anomalous dimensions is
of the form $\as^2/(N-1)$, and thus no stronger than
that at one loop. Naively one expects that at $l$ loops
$\gamma^{gg}_N\sim \as^{l}/(N-1)^{2l-1}$ because
at each order there is both an extra mass singularity and an extra
collinear singularity,\cite\Gross\ which would wreck double scaling by
inducing a strong power-like growth. However many of the
singularities cancel,\cites{\BFKL,\CCFM} so that the true
behaviour at $l$ loops is $\gamma^{gg}_N\sim(\as/(N-1))^{l}$.
The (scheme independent) coefficients of
these remaining singularities can be calculated\cite\Jaro\ and in fact turn out
to be very small; indeed the coefficients at $2$, $3$
and $5$ loops actually vanish. This means that although
the higher singularities may eventually produce a power-like growth
in the Regge limit like $x^{-\lambda_{\rm S}}$, where $\lambda_{\rm S} =
12\ln 2 \as/\pi$ is the radius of convergence of the
sum of singularities, no indication of such behaviour has yet been seen at
HERA.\cites{\das,\summing}

The behaviour of nonsinglet unpolarized parton distributions at small
$x$ is rather different, however. To begin with, the leading
singularity in $\gamma^{qq,{\rm NS}}_{N}$ is not at
$N=1$ but at $N=0$: all singularities at $N=1$ cancel. According
to Regge behaviour as
$x\to 0$ $q_{NS}(x,t_0)\sim x^{-\lambda}$, with $\lambda\simeq 0.5$,
which now counts as a `hard' boundary condition. So instead of the
double scaling behaviour \dasg, we now have
\eqn\hardbc{q_{NS}(x,t)\sim \tilde{\cal N}\exp(
{\lambda\xi+(\tilde\gamma^2/\lambda-\tilde\delta)\zeta}),}
for $\rho\equiv\sqrt{\xi/\zeta}\gsim\tilde\gamma/\lambda$,
$\tilde\gamma$ and $\tilde\delta$ constants. It follows
immediately that at small $x$ $q_{NS}$ is considerably smaller
$q_{S}$, and thus $F_2^p-F_2^n$ will be much more difficult to measure
than $F_2^p$ alone.
The two loop correction is now rather larger since
to $O(\as^2)$ $\gamma^{qq,{\rm NS}}_{N}$ behave as $\as^2/N^3$ as $N\to 0$,
and indeed at $l$ loops there is now no
cancellation of double logarithmic singularities,\cite\KiLi\ so
$\gamma^{qq,{\rm NS}}_{N}\sim\alpha^{l}/N^{2l-1}$. Again the
coefficients of the singularities can be computed,\cite\EMR\ and turn
out to be large.\footnote{$^{\rm a}$}{\footnotefont\baselineskip=8pt
Although there is now no factorization theorem to gaurantee that the
resulting leading singularity anomalous dimension is scheme
independent, the results at two loops (both unpolarized and
polarized\cite\MevN) are correct,\cite\BlVo\ and the
behaviour of the \MS\ coefficient functions
at $O(\as)$ and\cite\ZivN\ $O(\as^2)$ is not so singular as to introduce
scheme dependence into the anomalous dimension (at least at NNLO). It thus
seems that the implications of ref.\xcite\KiLi\ deserve to be
taken seriously.} It follows that even if
$q_{NS}(x,t_0)$ had been soft, asymptotically $q_{NS}(x,t)$ will
eventually grow as $x^{-\lambda_{\rm NS}}$, where\cite\KiLi\
$\lambda_{\rm NS}=\sqrt{8\as/3\pi}\sim\half$. As it is, this
behaviour should set in very quickly, and indeed it seems to be in
agreement with existing NMC and CCFR data.

Polarized distributions, both nonsinglet and singlet, have a soft
Regge behaviour:\cites{\Heim,\Kuti}
$\Delta q_{NS}(x,t_0)\sim \Delta q_{S}(x,t_0)
\sim x^{-\lambda}$, with $-0.5\lsim\lambda\lsim 0$ while again all
perturbative singularities at
$N=1$ cancel. Indeed at LO all anomalous dimensions behave as $\as/N$,
and thus one might expect a double scaling growth for both nonsinglet
and singlet distributions:\cites{\BFR,\AhRo-\pg}
\eqn\sxxNLO{\eqalign{\Delta q_{\rm NS}(x,t)&\sim{\cal N}_{\rm NS}
\sigma^{-1/2}\exp({2\gamma_{\rm NS}\sigma-\delta_{\rm NS}\zeta}),\cr
v_\pm(x,t)&\sim{\cal N}_\pm
\sigma^{-1/2}\exp({2\gamma_{\pm}\sigma-\delta_\pm\zeta}),\cr}}
where $\gamma_{\rm NS}$, $\gamma_\pm$, $\delta_{\rm NS}$ and
$\delta_\pm$ are all (known) constants, and $v_\pm(x,t)$
are eigenvectors of the evolution: $v_\pm=(\Delta q_S^\pm,\Delta g^\pm)$,
$\Delta q_S^\pm= -c_\pm\Delta g^\pm$, with $c_\pm$ both positive
constants. Thus rather than the gluon driving the singlet quark, as
happens in the unpolarized case, here both gluon and singlet quark
grow together, but with opposite signs. It follows that in general
while $\Delta q_{\rm NS}$
and $\Delta g$ become large and positive as $x\to 0$ and $t\to\infty$,
$\Delta q_{\rm S}$ becomes large and negative, and $g_1(x,Q^2)$
can then exhibit strong fluctuations due to interference between the
various contributions.\cite\BFR\ Furthermore $g_1^p$ and $g_1^n$ can behave
rather differently, since $\Delta q_{\rm NS}$ grows on the same
footing as $\Delta q_{\rm S}$.

However, these results must be
interpreted with great care since at small enough $x$ higher order
singularities can quickly become important. All the two loop anomalous
dimensions behave as $\as^2/N^3$, and although the two loop
corrections are still numerically small in the small $x$ region accessible
currently,\cite\pg\ and in particular leave the singlet eigenvectors
unchanged at NLO, at yet smaller $x$ they soon become comparable to
the LO term. Just as in the nonsinglet unpolarized case, there is no
evidence that the double logarithmic singularities cancel: indeed the
leading singularities in the nonsinglet polarized channel have already
been computed,\cite\KiLi\ and the resulting power behaviour of
$\Delta q_{\rm NS}(x,t)$ turns out to be even stronger\cite\BER\ than that of
$q_{\rm NS}(x,t)$. The same seems to be true in the polarized singlet
channel,\cite\FiKi\ though the matrix of anomalous dimensions is not
yet known.

To summarise, as $1/x$ and $Q^2$ increase $\vert g_1(x,Q^2)\vert$ grows
rapidly due to perturbative evolution, and may go negative if
$\Delta g$ is large enough. The precise details of the behaviour at
very small $x$ have yet to be calculated, but will probably involve
rather complicated fluctuations in the $(x,Q^2)$ plane. However what
is already clear is that the small $x$ contribution to the first
moment of $g_1$ may have a strong $Q^2$ dependence. Since the
scale dependence of the complete first moment \gfmao\ is
perturbatively rather weak, this would necessarily imply a
compensating $Q^2$ dependence at larger $x$ driven by $\Delta g$,
which may be visible in existing data.

\newsec{Polarized Gluons?}

\nref\ANR{G.~Altarelli, P.~Nason and G.~Ridolfi, \PL\vyp{B320}{1994}{152}.}
\nref\GRW{M.~Gluck, E.~Reya and W.~Vogelsang, \PL\vyp{B359}{1995}{201}.}

\nref\GeSt{T.~Gehrmann and W.J.~Stirling, \ZP\vyp{C65}{1995}{461}.}

Fixed target experiments with a fixed beam energy measure the polarization
asymmetries, and thus indirectly $g_1(x,Q^2)$, along a curve
$Q^2_{\rm exp}(x)$. For the SMC experiments\cites{\SMCp,\SMCd} this
curve goes from $(0.7,50\GeV^2)$ to $(0.003,1\GeV^2)$, while for the
E143 experiments the beam energy is lower, so the $Q^2$ is lower for
all $x$, the curve reaching from $(0.7,9\GeV^2)$ to $(0.03,1\GeV^2)$.
Combining both experiments thus gives us two values of $Q^2$ for each
value of $x$, making it possible to search for purely evolutionary
effects.

Several fits to the available data have been made, by evolving a
standard parameterization of $\Delta q_{\rm NS}(x,t_0)$,
$\Delta q_{\rm S}(x,t_0)$ and $\Delta g(x,t_0)$
from the starting scale (usually taken to be $Q_0^2=1\GeV^2$) up to the
data. Until recently the evolution was performed at
LO\cites{\ANR,\GRW} or in some hybrid `nLO'
approximation\cites{\GeSt,\BFR} in which
the one loop anomalous dimensions are used together with one loop
coefficient functions in some sensible scheme (one in which
the first moments are given by \cofm), in an effort to assess
the effect of the direct coupling of $\Delta g$ to $g_1$. The
calculation\cite\MevN\ of the two loop anomalous dimensions has
now made complete NLO calculations\cite\pg\ possible.

\topinsert
\vskip-2.5truecm
\vbox{\hbox{\hskip-1truecm
\hfil\epsfxsize=7truecm\epsfbox{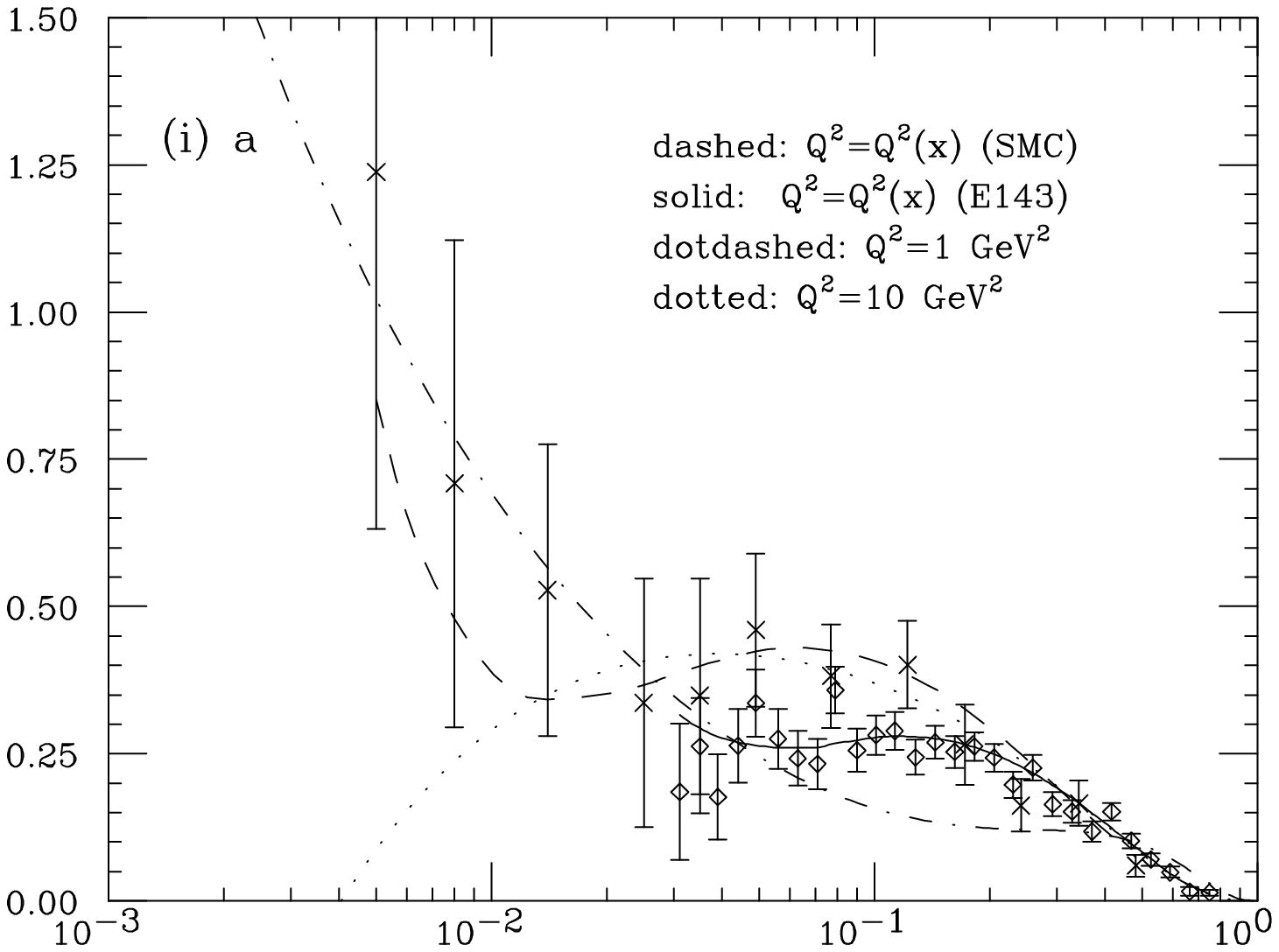}
\hskip-0.5truecm
\epsfxsize=7truecm\epsfbox{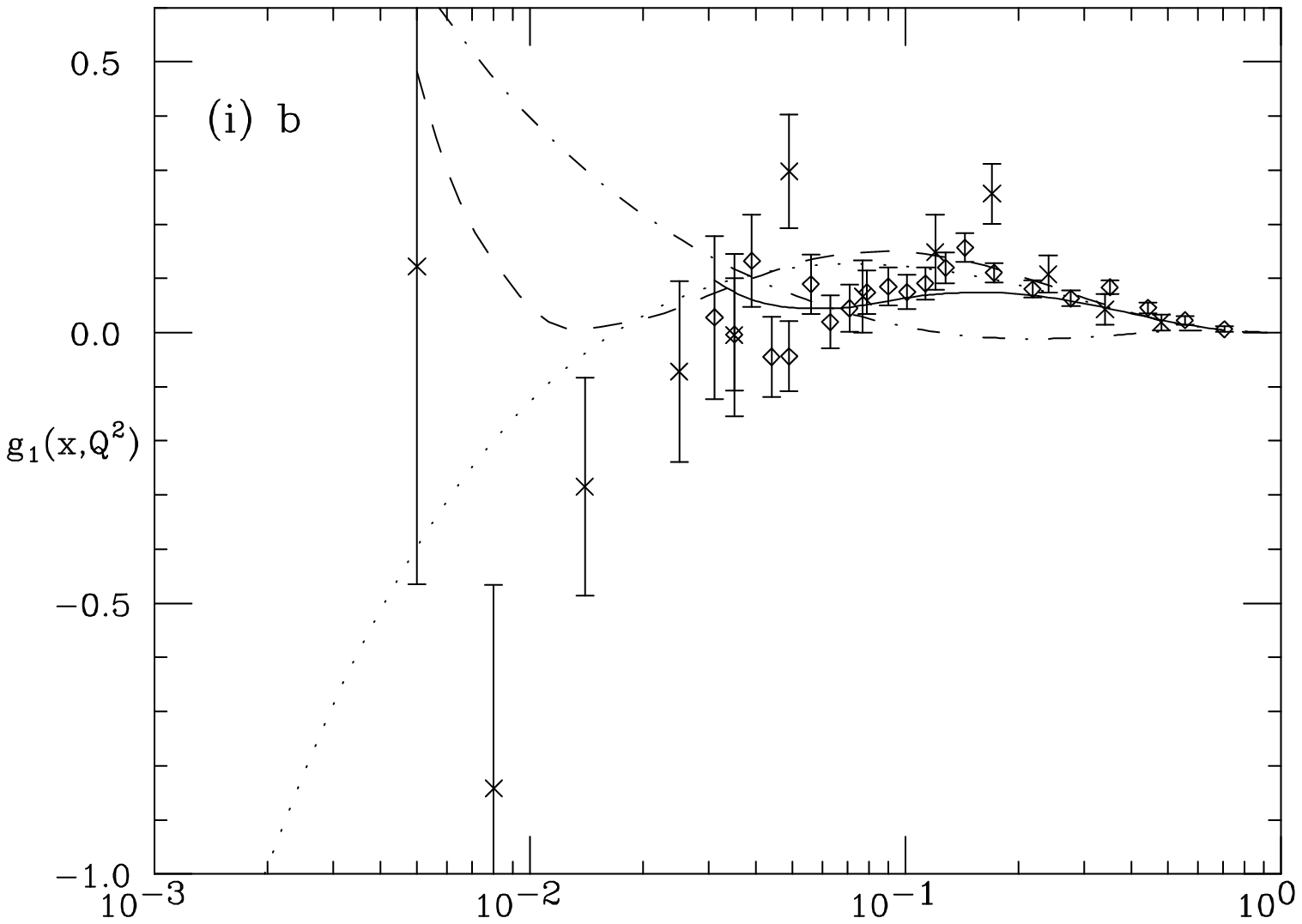}\hfil}
\vskip-4.5truecm
\hbox{\hskip-1truecm
\hfil\epsfxsize=7truecm\epsfbox{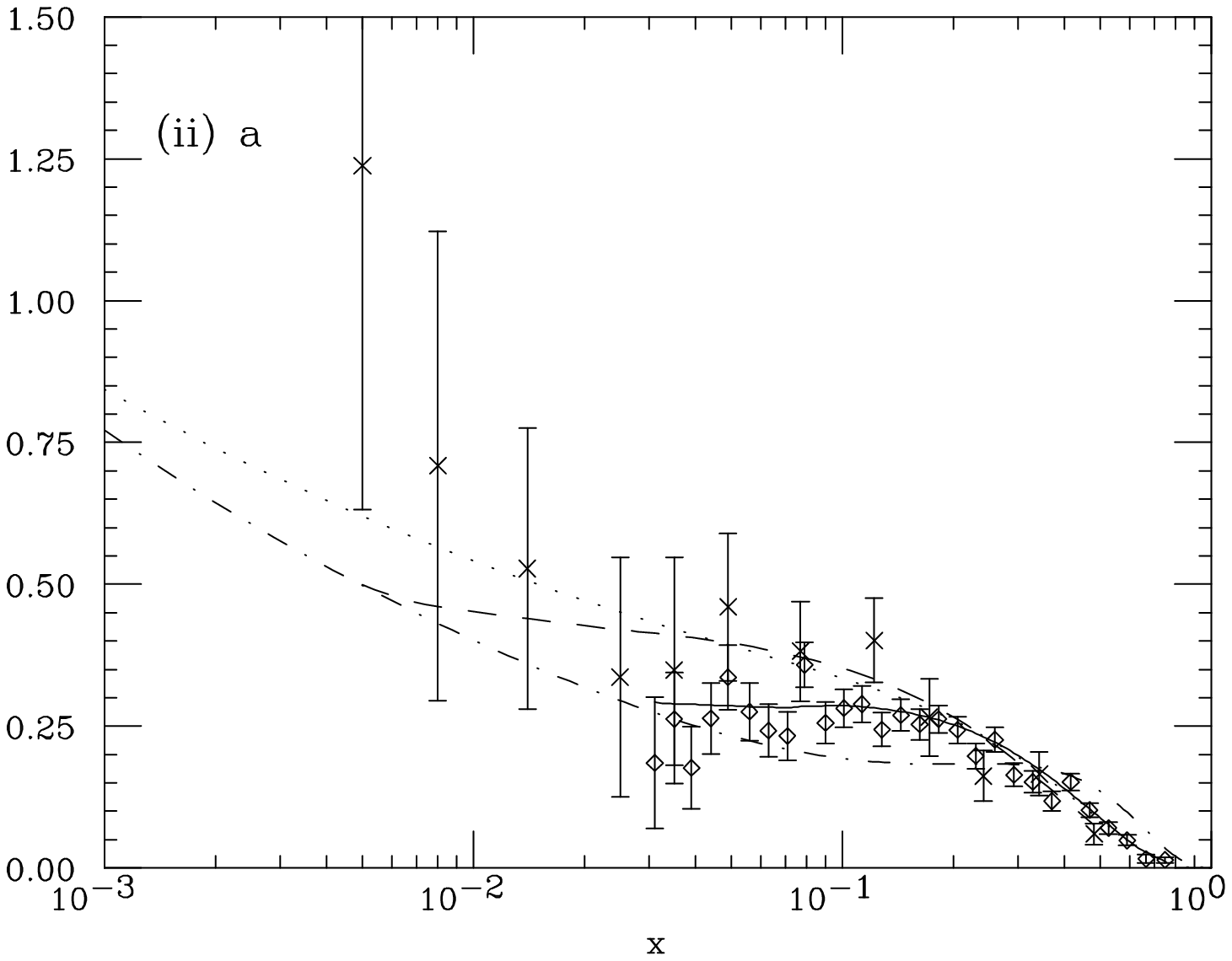}
\hskip-0.5truecm
\epsfxsize=7truecm\epsfbox{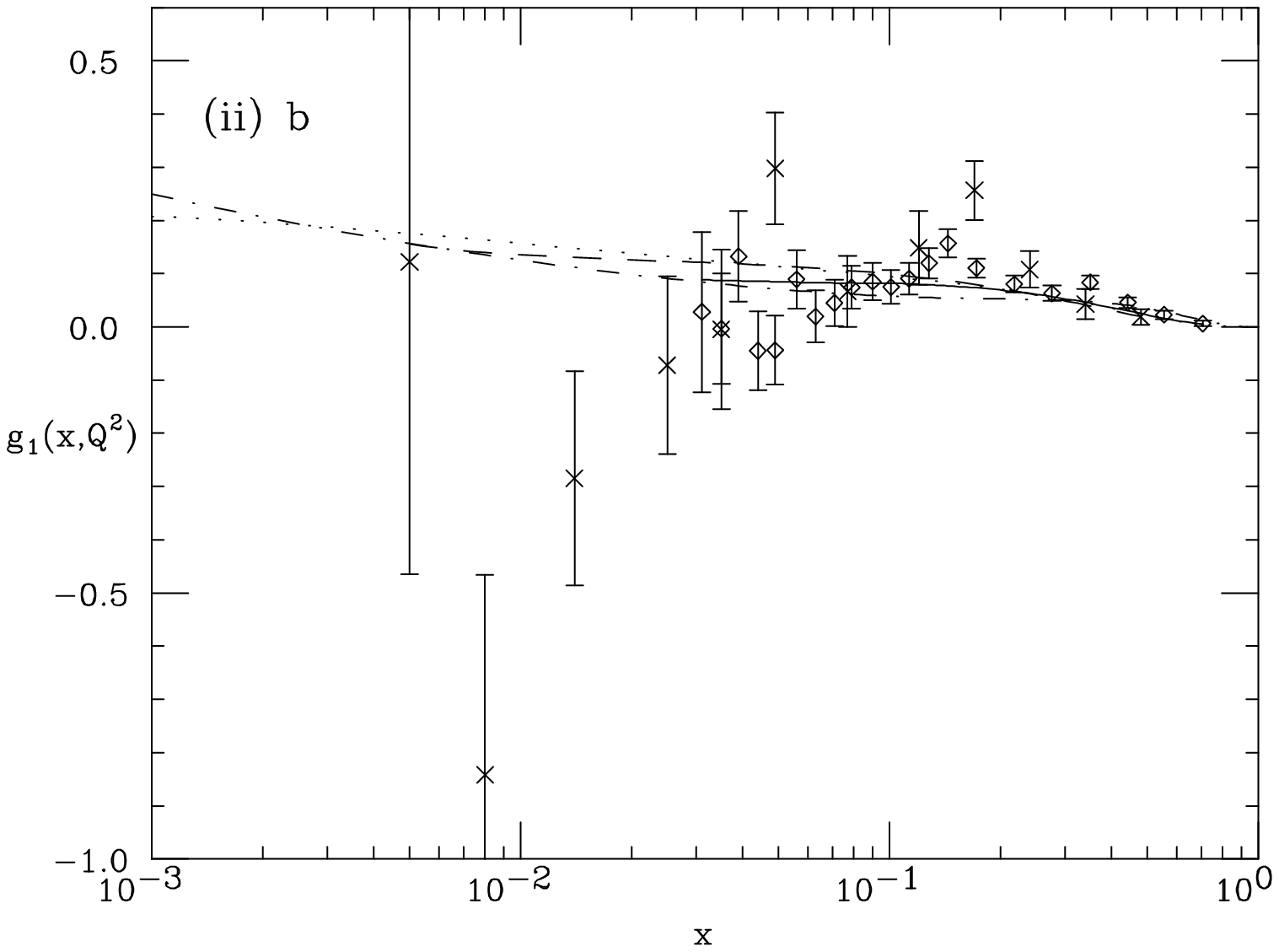}\hfil}
\vskip-3truecm
\bigskip\noindent{\footnotefont\baselineskip6pt\narrower
Figure 1: The SMC\cites{\SMCp,\SMCd} (crosses) and E143\cites{\SLACp,\SLACd}
(diamonds) data for (a) $g_1^p$ and (b) $g_1^d$ plotted against $x$.
The curves correspond
to (i) maximal gluon and (ii) minimal gluon fits.\cite\BFR}}
\medskip
\endinsert

\topinsert
\vskip-2.5truecm
\vbox{\hbox{\hskip-1truecm
\hfil\epsfxsize=7truecm\epsfbox{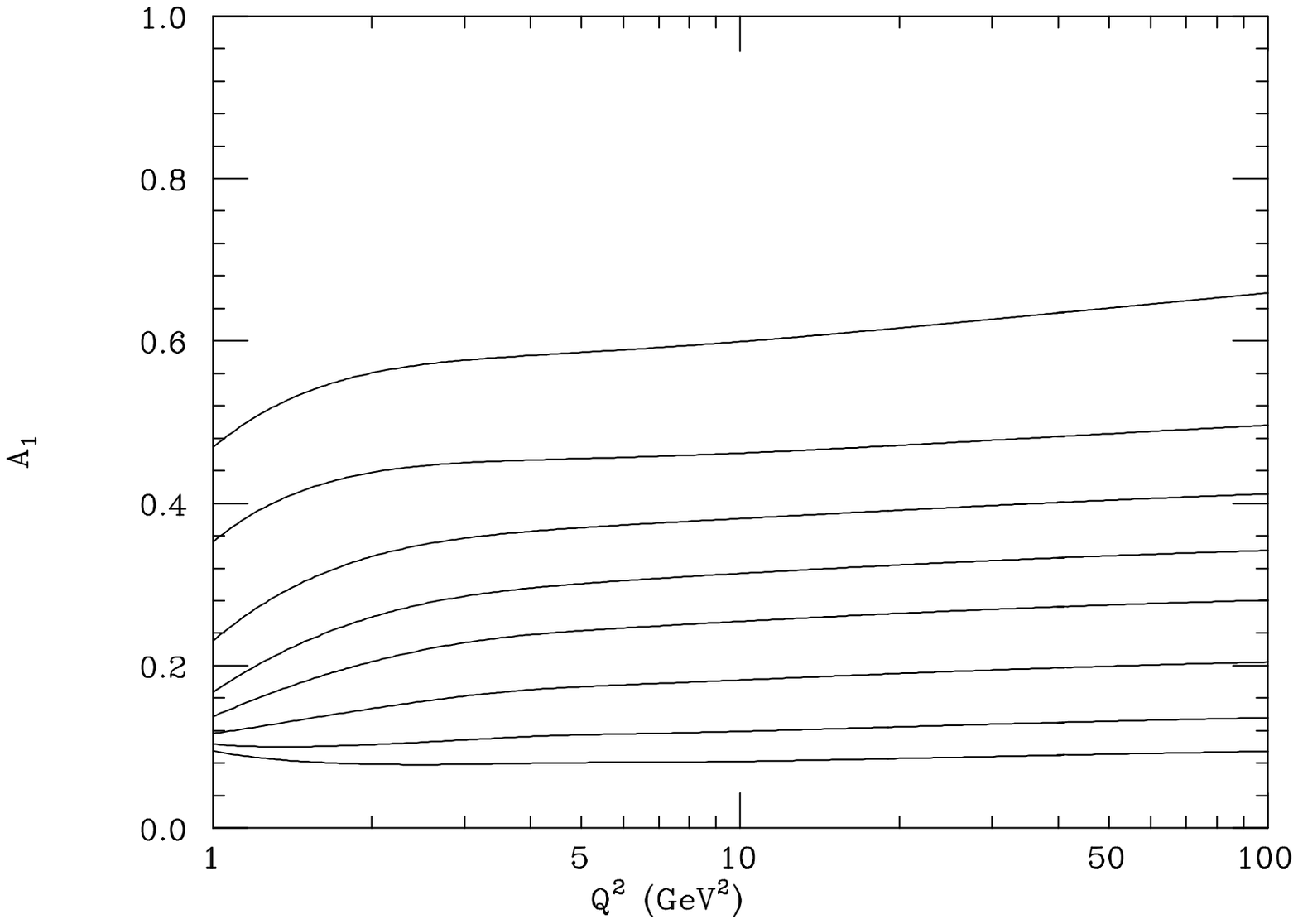}
\hskip-0.5truecm
\epsfxsize=7truecm\epsfbox{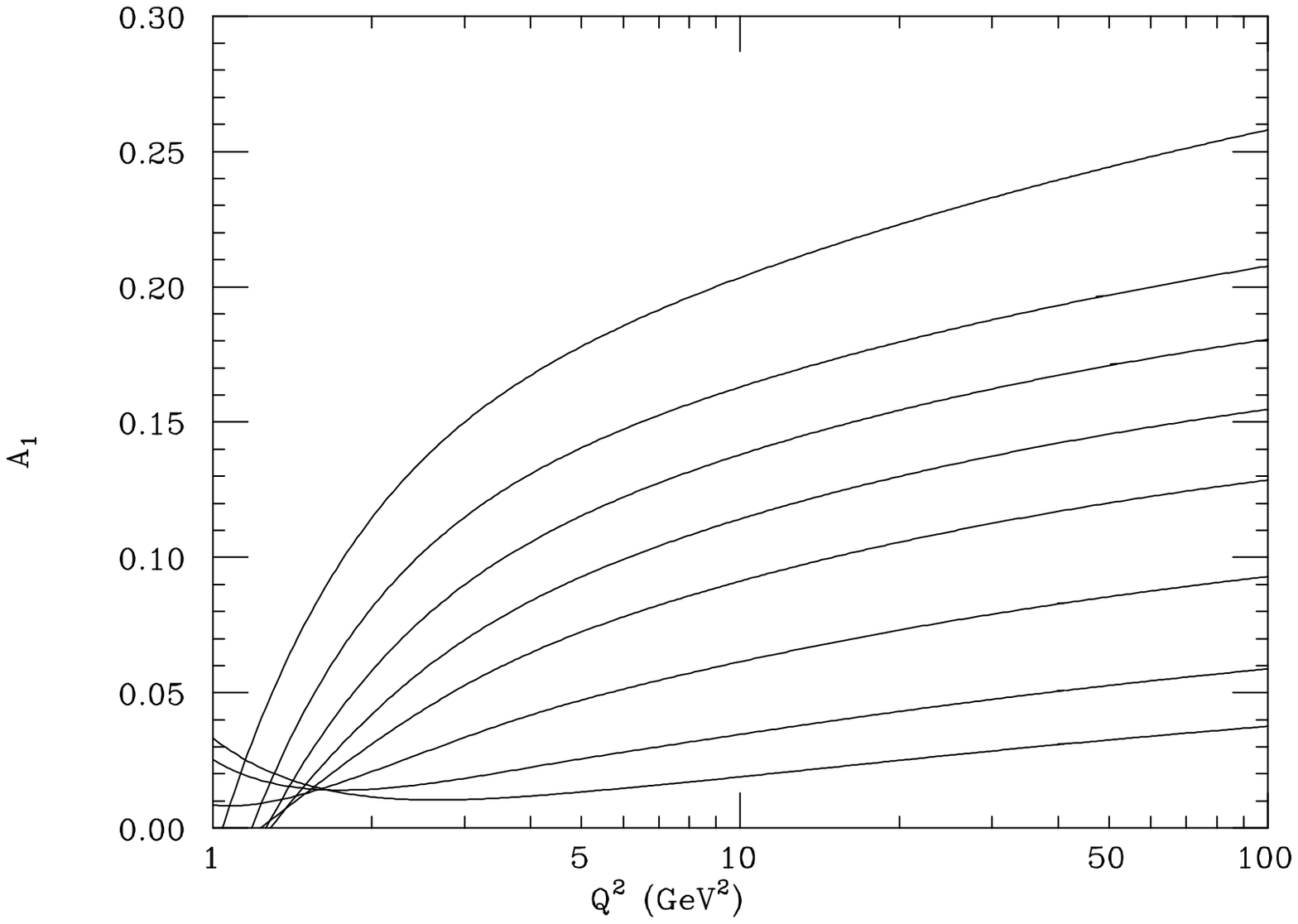}\hfil}
\vskip-4.5truecm
\hbox{\hskip-1truecm
\hfil\epsfxsize=7truecm\epsfbox{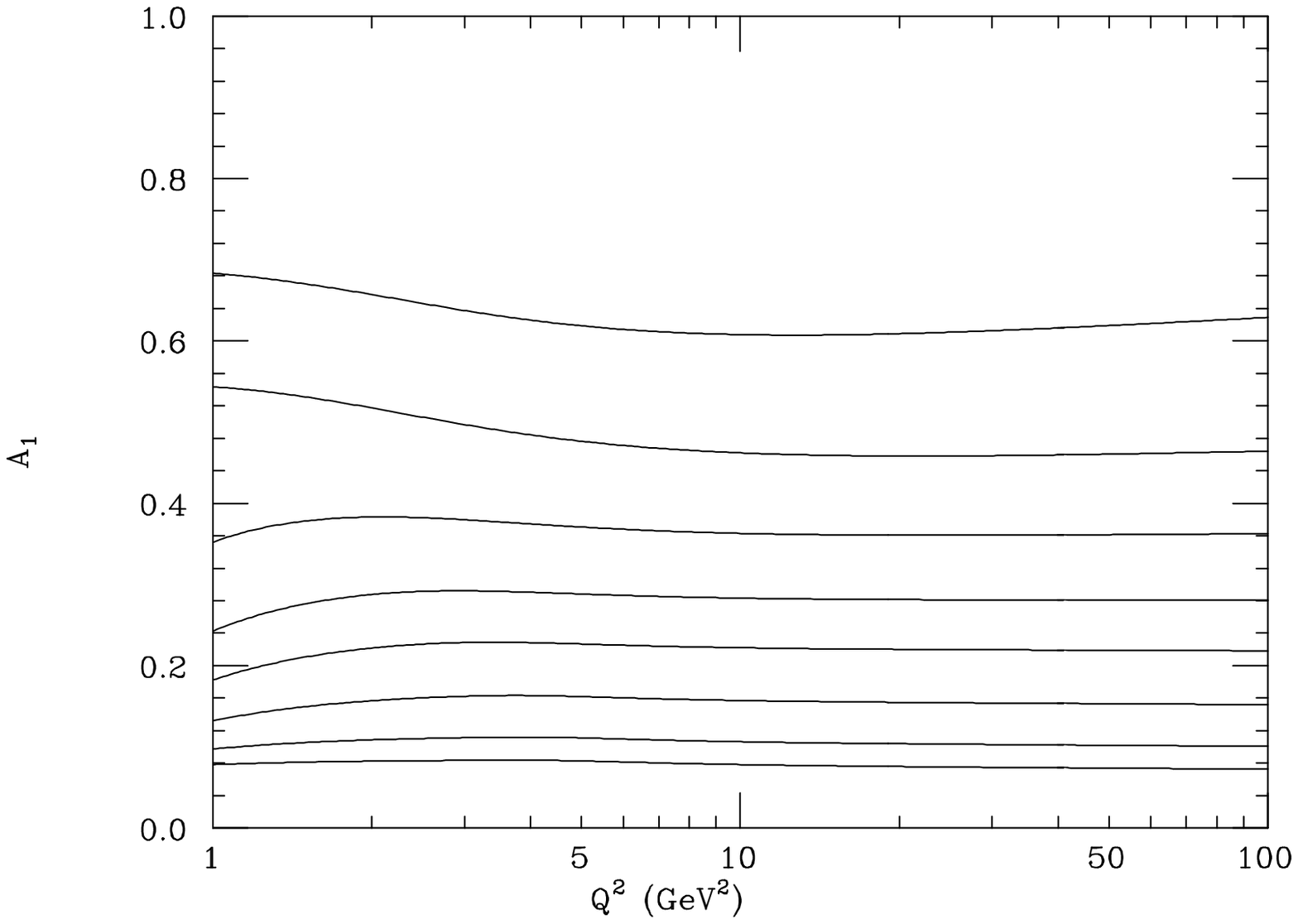}
\hskip-0.5truecm
\epsfxsize=7truecm\epsfbox{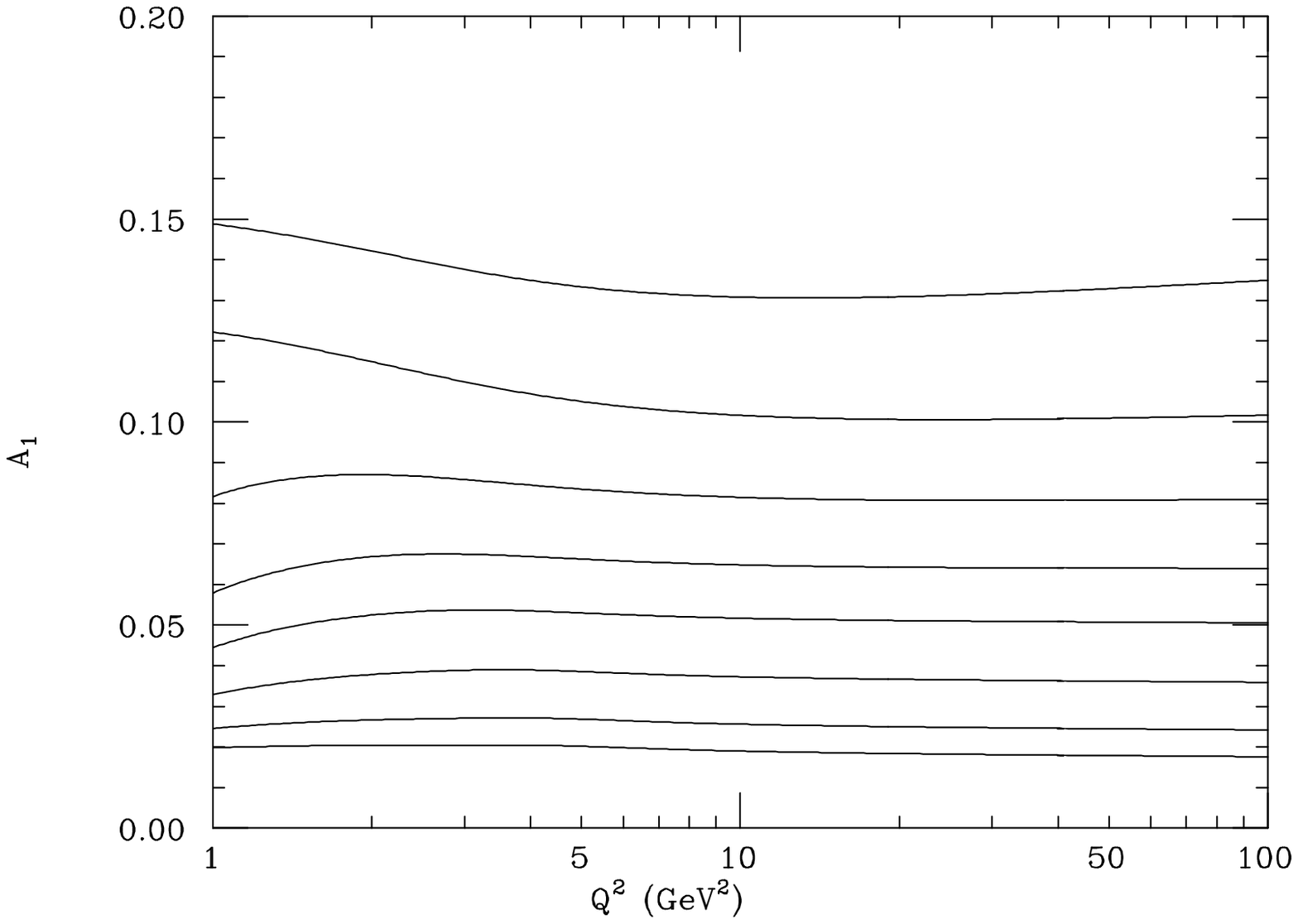}\hfil}
\vskip-3truecm
\bigskip\noindent{\footnotefont\baselineskip6pt\narrower
Figure 2: The asymmetries corresponding to the four fits in fig.~1,
plotted against $Q^2$.
{}From top to bottom the curves correspond to
$x=0.5,0.35,0.25,0.175,0.125,0.08,0.05,0.035$.}}
\medskip
\endinsert

At LO the effects of evolution are generally small, and in particular
the asymmetries are reasonably scale independent.\cite\ANR\ However at
nLO the direct gluon coupling can have a dramatic effect,
especially if $\Delta g$
is large. In fig.~1a we show the result of two nLO fits\cite\BFR\ to
the proton data\cites{\SMCp,\SLACp} alone, one (the `minimal gluon')
in which $\Delta g_1(t_0)=0$, while $\Delta q^{\rm S}_1(t_0)$ is
fitted, the other (the `maximal gluon') in which
$\Delta q^{\rm S}_1(t_0)=a_8$ (as
expected from the Zweig rule) while $\Delta g_1(t_0)=0$ is fitted.\footnote
{$^{\rm b}$}{\footnotefont\baselineskip=8pt In both cases
$\Delta q^{\rm S}_1$ is fixed by eqn.\nsaxch. Other details of the
fits may be found in ref.\xcite\BFR.} The minimal gluon fit shows that there
is considerable evolution from the E143 data up to the SMC data, but
below the crossover (at $x\sim 0.3$) this is fairly uniform in $x$.
The maximal gluon fit is much more dramatic: there is a second
crossover at $x\sim 0.03$ while at yet smaller $x$ $g_1^p(x,Q^2)$ becomes
increasingly negative as $Q^2$ increases. The evolution between the
two crossovers is then correspondingly larger. This is particularly
evident in the corresponding asymmetries (fig.2a), which are fairly
flat for the minimal gluon but rise quite steeply with $Q^2$ for the
maximal gluon.

Although on the basis of these fits to the proton data alone it was
not possible to distinguish between minimal and maximal gluon, the
corresponding predictions fig~1b for the deuteron structure function
are more distinct, essentially because the deuteron is predominantly
singlet, and a large direct gluonic contribution to $g_1$ thus tends to
make $g_1^d$ negative at small $x$. The data\cites{\SMCd,\SLACd} seem
to prefer a maximal gluon. The strong growth in the deuteron
asymmetry which this generates is apparent in fig.~2b.

This conclusion has recently been confirmed by a complete NLO
calculation, with both proton and deuteron data\cites{\SMCp-\SLACd}
included in the fit.\cite\pg\ The parton distributions turn out to be
surprisingly well determined, with $\Delta q^{\rm S}_1(t_0)=0.5\pm 0.1$,
$\Delta g_1(t_0)=1.5\pm 0.8$. The Zweig rule expectation
$\Delta s\simeq 0$ is thus confirmed experimentally, while
the discrepancy in the Ellis-Jaffe sum rule\cite\EJ\ is accounted for
almost entirely by
a direct polarized gluon contribution, just as was conjectured in
ref.\xcite\AlRo. The behaviour of the parton distributions at large $x$
is roughly consistent with quark
counting rules. At small $x$ the nonsinglet distribution is
singular (behaving as $x^{-\lambda}$, $\lambda=0.7\pm 0.2$) while the
singlet quark and gluon distributions are generally either flat or
valencelike, as expected from the theoretical considerations reviewed
in the previous section.

\topinsert
\vskip-2.5truecm
\vbox{\hbox{\hskip-1truecm
\hfil\epsfxsize=7truecm\epsfbox{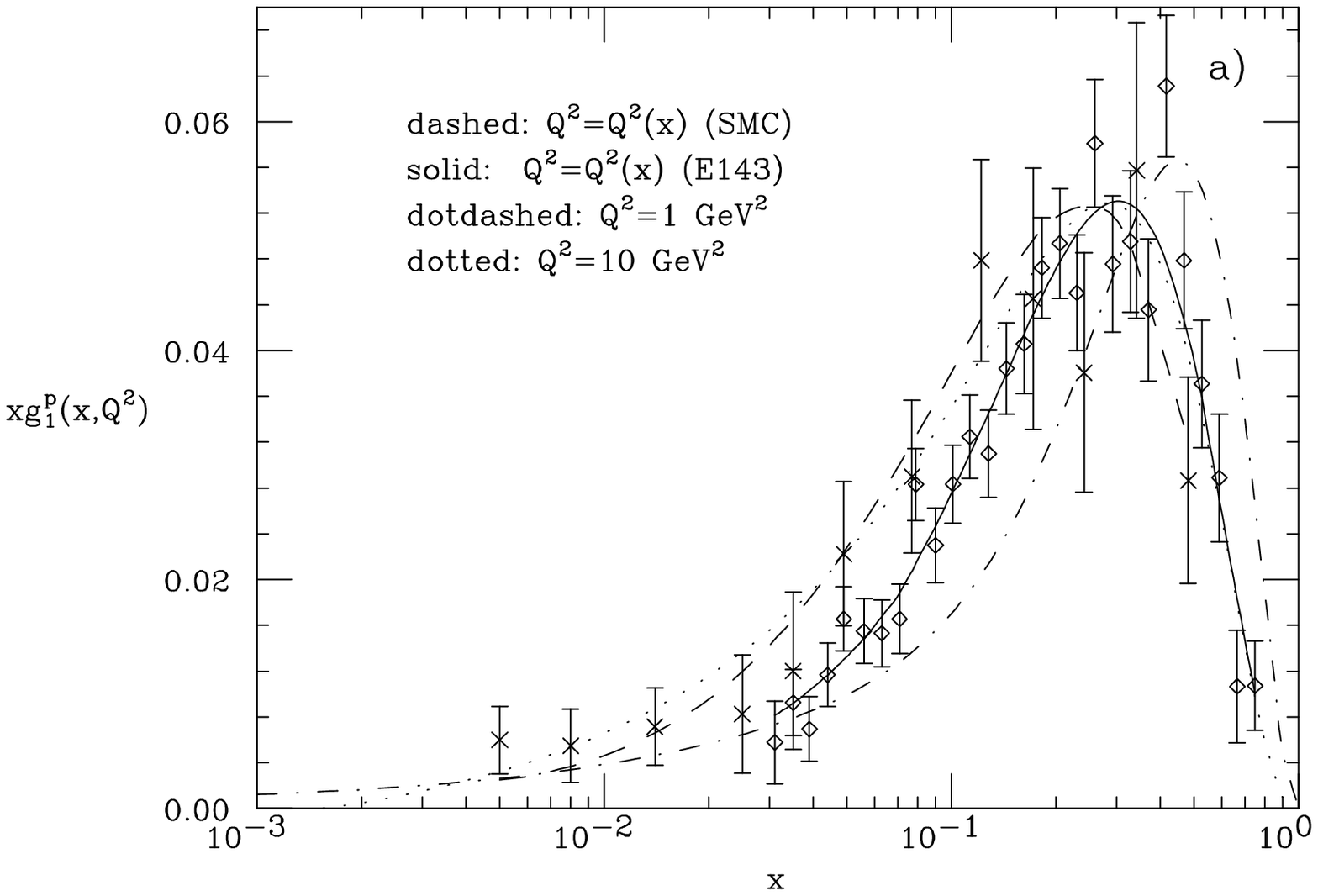}
\hskip-0.5truecm
\epsfxsize=7truecm\epsfbox{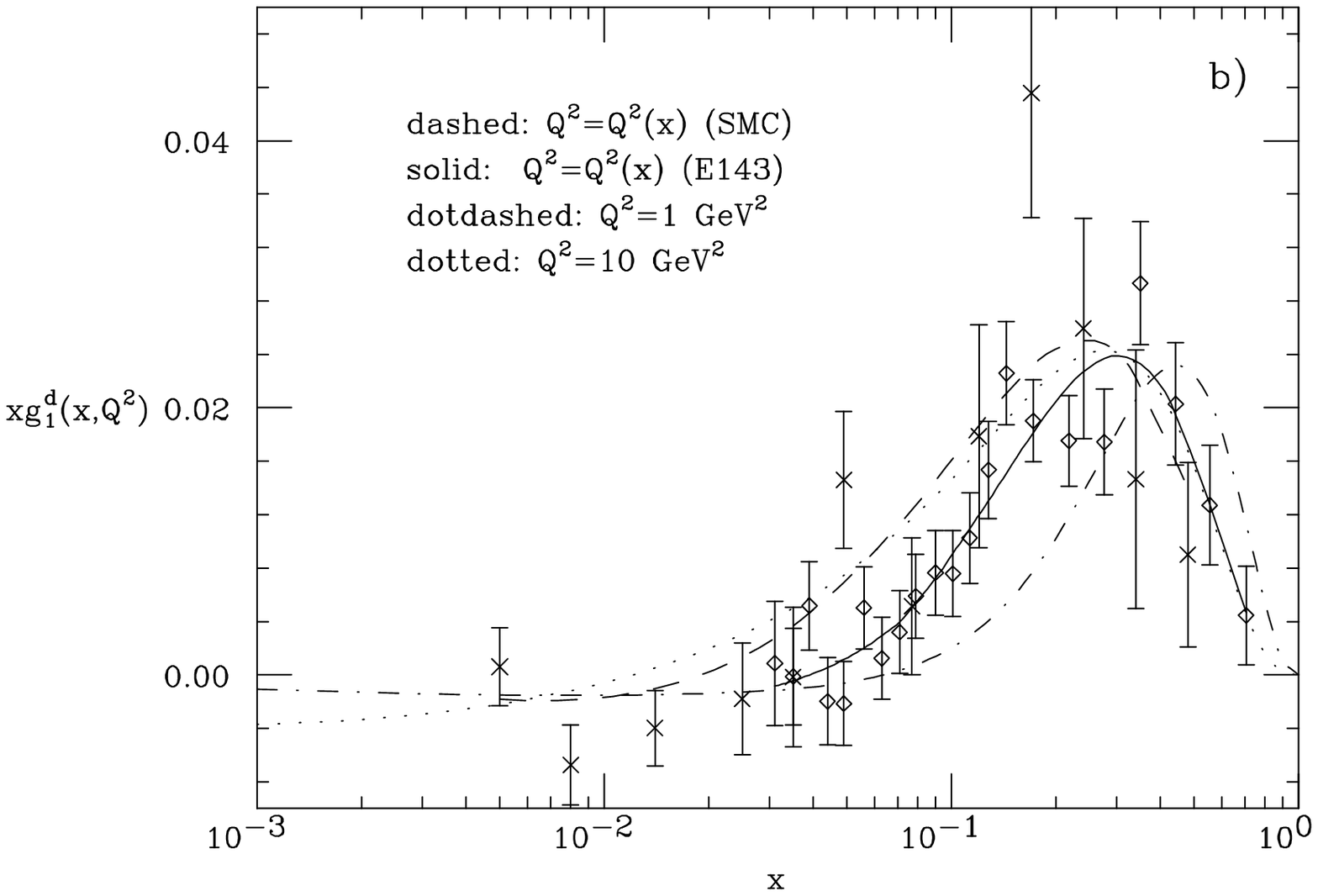}\hfil}
\vskip-3truecm
\bigskip\noindent{\footnotefont\baselineskip6pt\narrower
Figure 3: The structure functions $xg_1^p(x,Q^2)$ and $xg_1^d(x,Q^2)$
plotted against $x$. The notation of the data
points\cites{\SMCp-\SLACd} and fitted curves\cite\pg\ is the same as
in fig.~1.}}
\medskip
\endinsert

Using this fit (displayed in fig.~3) to determine the first
moments $\Gamma_1^p$,
$\Gamma_1^d$ and $a_0$ (as defined in eqn.\evecs) we find\cite\pg
\eqn\fm{\eqalign{
\Gamma_1^p&=0.122\pm 0.013\>\hbox{(exp.)}
\epm{0.011}{0.005}\>\hbox{(th.)},\cr
\Gamma_1^d&=0.025\pm 0.013\>\hbox{(exp.)}
\epm{0.012}{0.004}\>\hbox{(th.)},\cr
a_0&=0.14\pm 0.10\>\hbox{(exp.)}
\epm{0.12}{0.05}\>\hbox{(th.)},\cr}}
at $Q^2=10\GeV^2$. The central values are lower than those given by
the experimental collaborations because of the scale dependence both of the
asymmetries in the measured region and of the small $x$
extrapolations: the experimental
uncertainty is larger because of the uncertainty in the size of the
polarized gluon distribution which drives this evolution. There
is also a theoretical error
which is predominantly due to an estimate of NNLO corrections: these
can be large because the two loop coupling of $\Delta g_1$ to
$\Gamma_1$ is effectively NLO (compare \gfmao\ with \gonemom\ when $N=1$).

\newsec{Desiderata}

\nref\BaHu{D.P.~Barber and V.~Hughes, in these proceedings.}

In conclusion, the effects of perturbative evolution have to be taken
into account when extracting axial charges from polarization
asymmetries, since they may be large due to the anomalous coupling to
polarized gluons. Conversely, the structure of the evolution seen in
the combined proton and deuteron data sets indicates that the
polarized gluon distribution may indeed be large, making a substantial
contribution to the nucleon spin. In order to make this conclusion
more definite, more data over a range of $Q^2$ at moderate values of
$x$ from one experiment are needed. Polarization of the
protons at HERA,\cite\BaHu\ enabling a measurement of $A_1^p$ (and perhaps
eventually $A_1^d$)  would further pin down the small $x$ contribution
and the size of the gluon. Data for moderately small $x$ (say
$0.001\lsim x\lsim 0.01$) but over a wide range of $Q^2$ would be most
useful. Indeed the behaviour of polarized structure
functions at small $x$ promises to be a very rich subject, for both
theorists and experimentalists alike.

\appendix{Acknowledgments}
All of the original work described in this article was done in collaboration
with Stefano Forte and Giovanni Ridolfi. I would also like to thank Stefano
Forte for a critical reading of the manuscript, and Bernard
Frois and Vernon Hughes for organising this school in such a beautiful place.

\immediate\closeout\rfile\writestoppt
\bigskip
\noindent{{\bf References}}\smallskip{\frenchspacing%
\parindent=20pt
\ninepoint\baselineskip=11pt
\escapechar=` \input refs.tmp\vfill\eject}\nonfrenchspacing

\end